\documentclass[10pt]{article}

\usepackage{amsmath}
\usepackage{amssymb}

\usepackage{graphicx}

\usepackage{cite}

\usepackage{color}

\usepackage{booktabs}

\usepackage[labelfont=bf,labelsep=period,justification=raggedright]{caption}

\makeatletter
\renewcommand{\@biblabel}[1]{\quad#1.}
\makeatother

\date{}

\begin{document}

\begin{flushleft}
{\Large
\textbf{Bayesian Analysis for miRNA and mRNA Interactions Using Expression Data}
}
\\
Mingjun Zhong,
Rong Liu,
Bo Liu
\\
\bf Department of Biomedical Engineering, Dalian University of Technology, Dalian, P.R. China
\\
E-mail: zhong@dlut.edu.cn, mingjun.zhong@gmail.com
\end{flushleft}

\section*{Abstract}
MicroRNAs (miRNAs) are small RNA molecules composed of 19--22 nt, which play important regulatory roles 
in post-transcriptional gene regulation by inhibiting the translation of the mRNA into proteins or 
otherwise cleaving the target mRNA. Inferring miRNA targets provides useful information for understanding 
the roles of miRNA in biological processes that are potentially involved in complex diseases. 
Statistical methodologies for point estimation, such as the Least Absolute Shrinkage and 
Selection Operator (LASSO) algorithm, have been proposed to identify the interactions of miRNA and mRNA 
based on sequence and expression data. In this paper, we propose using the Bayesian LASSO (BLASSO) and 
the non-negative Bayesian LASSO (nBLASSO) to analyse the interactions between miRNA and mRNA using expression data. 
The proposed Bayesian methods explore the posterior distributions for those parameters required to model 
the miRNA--mRNA interactions. These approaches can be used to observe the inferred effects of the miRNAs 
on the targets by plotting the posterior distributions of those parameters. For comparison purposes, 
the Least Squares Regression (LSR), Ridge Regression (RR), LASSO, non-negative LASSO (nLASSO), 
and the proposed Bayesian approaches were applied to four public datasets.
We concluded that nLASSO and nBLASSO perform best in terms of sensitivity and specificity. 
Compared to the point estimate algorithms, which only provide single estimates for those parameters, 
the Bayesian methods are more meaningful and provide credible intervals, which take into account 
the uncertainty of the inferred interactions of the miRNA and mRNA. Furthermore, 
Bayesian methods naturally provide statistical significance to select convincing inferred interactions, 
while point estimate algorithms require a manually chosen threshold, which is less meaningful, 
to choose the possible interactions. A Matlab implementation of the algorithms described in 
the present paper is available at http://code.google.com/p/nblasso/.


\section*{Introduction}
Mature miRNAs, a large family of regulatory RNAs expressed in animals and plants, are processed from primary transcripts in two steps by enzymes of the RNase III endonucleases Drosha in the nucleus and Dicer in the cytoplasm \cite{Lee03,Hutvagner05}. In the cytoplasm, miRNA is separated into two single strands by helicase enzymes. One strand is combined with an Argonaute protein to form the RNA-induced silencing complex (RISC). The RISC then mediates the regulation of gene expression by base pairing imperfectly to the 3'-untranslated region of the target mRNAs to inhibit protein synthesis \cite{Carthew09}.\\


Recent research has revealed that there are more than 1400 human miRNA sequences \cite{Kozomara11}. MicroRNA is reported to play important roles in various regulation processes, including proliferation, apoptosis, differentiation and development, cellular identity, and pathogen--host interactions \cite{Parker07,Pillai07,Carthew09}. Given the measured sequences of miRNA and mRNA, an important task is to identify the regulatory interactions between them. Experimental and sequence-based methods such as miRanda \cite{Betel08}, TargetScan \cite{Lewis05}, miRBase \cite{Griffiths08} and PicTar \cite{Krek05} have been proposed to predict the targets of miRNA regulations. These algorithms are mainly based on the experimentally determined rules of miRNA regulation. The most general feature of miRNA regulation is that the 3'-untranslated region of the target mRNAs is complementary to the seed region (nucleotides 2--7) of the miRNA \cite{Lewis03}. It has been shown that computational approaches to combining miRNA and mRNA high-throughput expression data with sequence-based putative predictions can greatly reduce the number of potential targets of miRNA regulation \cite{Muniategui12b}.\\

Various computational algorithms have been proposed to identify the mRNA and miRNA interactions using high-throughput 
expression datasets. These algorithms include correlation, mutual information, mutiple linear regression, 
partial least squares, the Least Absolute Shrinkage and Selection Operator (LASSO) \cite{Tib96}, 
non-negative LASSO (nLASSO or the TaLASSO as in \cite{Muniategui12a}), 
variational approximations for linear regression models (GenMiR++), HCTarget, the Bayesian graphical method \cite{Stingo10}, etc. 
A review of these algorithms can be found in \cite{Muniategui12b}. Among these methods, 
GenMiR++ \cite{Huang07}, HCTarget \cite{Su11} and the Bayesian graphical approach \cite{Stingo10} 
are probabilistic approaches. 
HCTarget and the Bayesian graphical method employ Markov chain Monte Carlo (MCMC) schemes 
to sample the posterior densities for the required parameters, while the GenMiR++ employs 
a variational approach to approximate the densities of the parameters. 
The graphical Bayesian method of \cite{Stingo10} is methodologically different from the regression-based approaches and is not directly comparable with the Bayesian methods in this paper. Both LASSO and nLASSO have been applied to linear regression models to model miRNA--mRNA interactions \cite{Lu11,Muniategui12a}. In this paper, we propose to examine the efficiency of Bayesian LASSO (BLASSO) \cite{Park08} and non-negative Bayesian LASSO (nBLASSO) in this task.
It should be noted that one advantage to employing the Bayesian approaches is that they can provide the uncertainty of the inferred parameters, while the point estimate based algorithms can only estimate single points for the parameters. Like HCTarget and GenMiR++, BLASSO has no non-negative constraints on the linear regression model parameters. Since these non-negative constraints have been shown to be meaningful for identification of down-regulation targets of miRNA on the mRNA expressions \cite{Muniategui12a}, we propose an nBLASSO algorithm based on TaLASSO.  The non-negative constraint is employed to assume that miRNAs down-regulate their corresponding mRNA targets. In this study, LSR, RR, LASSO, nLASSO, BLASSO and nBLASSO were applied to the experimentally validated interactions of four expression datasets. Of the point estimate algorithms, LSR and RR produced similar results, with noisy representations that require thresholding, while LASSO and nLASSO were able to produce sparse representations for the mRNA--miRNA interactions. Since both the BLASSO and nBLASSO are unable to select variables automatically, we propose calculating the credible intervals to take into account the prediction uncertainties for the variable selections. The statistical significance is calculated for the credible intervals and then the variables are selected based on this statistical significance given a significance probability $\alpha$.

\section*{Materials and Methods}
\begin{figure}[!ht]
  \centering
  \includegraphics[width=\textwidth]{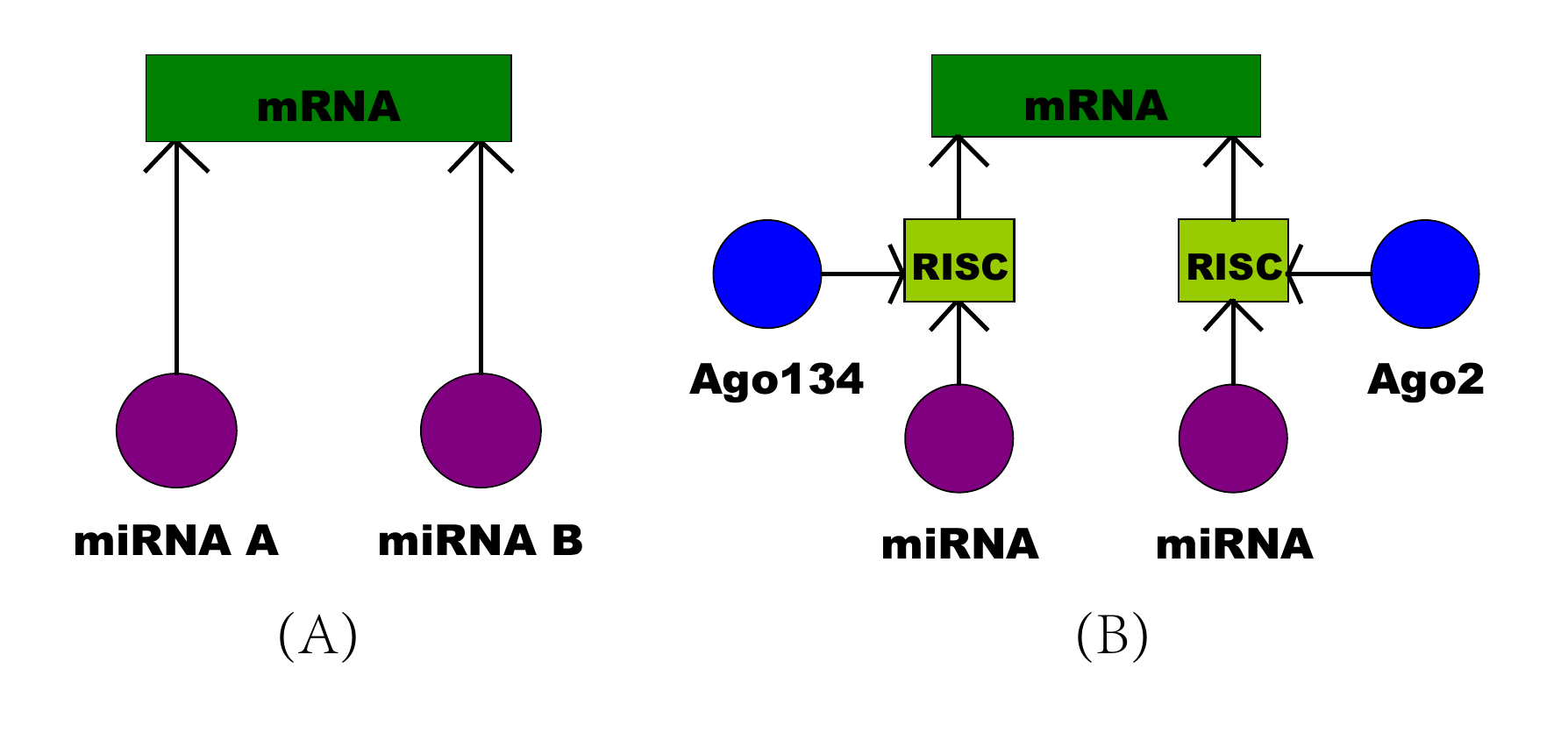}
  \caption{{\bf The models for mRNA and miRNA interactions}: (A) miRNA directly targets mRNA and (B) the complex RISC composed of miRNA and Argonaute proteins targets mRNA. }
  \label{networkAB}
\end{figure}

Given a miRNA and mRNA expression dataset, the interactions between the mRNA and miRNA are represented by the model
\begin{equation}
\mathbf{Y} = \mathbf{f}(\mathbf{X}) + \mathbf{E}
\end{equation}
where $\mathbf{Y}\in\mathcal{R}^{N\times P}$ represents the collections of the mRNA expression data with $N$ tissue samples and $P$ mRNAs, $\mathbf{f}$ is the function for modelling the interactions between $\mathbf{X}$ and $\mathbf{Y}$, $\mathbf{X}\in\mathcal{R}^{N\times M}$ represents the collections of the miRNA expression data with $M$ miRNAs, and $\mathbf{E}$ is the noise matrix. This is a general model for representing the miRNA targets and the function $\mathbf{f}$ could be non-linear. In this paper we only consider a linear model, which is then represented by
\begin{equation}
\mathbf{Y} = \mathbf{X}\mathbf{f} + \mathbf{E}
\end{equation}
where $\mathbf{f}\in\mathcal{R}^{M\times P}$ is a matrix. This model is similar to the bilinear model proposed in \cite{zhong11}, where both the $\mathbf{X}$ and $\mathbf{f}$ could be unknown or $\mathbf{X}$ is known and $\mathbf{f}$ is unknown. Since both the miRNA and mRNA expression data are observed, only the latter case is considered. An illustration of the model is shown in Figure \ref{networkAB}, where (A) models the direct interactions between the mRNA and miRNA and (B) models the interactions between the mRNA and the RISC, which is a complex composed of miRNA and Argonaute protein. Previous research \cite{Stanhope09,Lu11} has shown that model (B) has some advantages over model (A), but in this paper both models are considered. Therefore, $\mathbf{X}$ represents the miRNA for model (A) or the complex RISC for model (B). Note that $\mathbf{X}$ may be the multiplication of $-1$ and the miRNA expressions. Thus a positive value of $f_{mp}$ represents a negative regulation of the miRNA. Research has found that miRNA usually takes part in negative regulation in translational repression or target degradation and gene silencing \cite{Bartel09,Kus06}. However, miRNA may also be involved in positive regulation, for example in transcriptional and translational activations \cite{Vas07}. Thus, in order to model the roles of miRNA, we will consider both the cases $f_{mp}\in\mathcal{R}$ and $f_{mp}\geq0$.

\subsection*{Modelling both positive and negative regulation}
MicroRNAs can be involved in both positive and negative regulation of their targets. For one gene, we have $\mathbf{y}\in\mathcal{R}^{N\times 1}$ and $\boldsymbol{\beta}\in\mathcal{R}^{M\times 1}$, which is one column of $\mathbf{f}$. The model is represented by
\begin{equation}
\mathbf{y}=\mathbf{X\boldsymbol{\beta}}+\boldsymbol{\epsilon}
\end{equation}
where $\boldsymbol{\epsilon}$ is one column of the noise matrix $\mathbf{E}$. Assume that $\boldsymbol{\epsilon}\sim\mathcal{N}(0,\sigma^2\mathbf{I})$ is Gaussian noise with zero mean and variance $\sigma^2$. The main task is then to estimate the parameter vector $\boldsymbol{\beta}$ which models the miRNA targets. Several algorithms are available to resolve this problem, such as the LSR, RR, LASSO \cite{Hastie01} and BLASSO \cite{Park08}.
These algorithms are briefly described below, but for an in-depth review of their application to mRNA--miRNA interactions, see \cite{Muniategui12b}.

The least squares regression (LSR) for the linear model is defined by the optimization problem $\hat{\boldsymbol{\beta}}=argmin_{\boldsymbol{\beta}}\left\{\sum_{i=1}^N\left(y_i-\sum_{j}\beta_jx_{ij}\right)^2\right\}$.
The solution has the closed form $\hat{\boldsymbol{\beta}} = (\mathbf{X}^T\mathbf{X})^{-1}\mathbf{X}^T\mathbf{y}$.

On the other hand, the ridge regression (RR) solves the same optimization problem with the constraint $\sum_j\beta^2_j\leq t$. The solution is thus represented by
$\hat{\boldsymbol{\beta}} = (\mathbf{X}^T\mathbf{X}+\lambda\mathbf{I})^{-1}\mathbf{X}^T\mathbf{y}$
where $\lambda>0$ is a regulation parameter. In the particular case when the number of microRNAs is larger than the number of samples, i.e., $N<M$, the symmetric matrix $\mathbf{X}^T\mathbf{X}$ is not invertible, and thus the LSR is not applicable. However, RR could be applicable since the regulation parameter could be tuned to make the matrix $\mathbf{X}^T\mathbf{X}+\lambda\mathbf{I}$ be invertible. Note that the regulation parameter $\lambda$ and $t$ perform the same role and are inversely proportional. That is, a larger value for the parameter $\lambda$ (or a smaller value for $t$) increases the shrinkage of the regression coefficients.

The LASSO estimate for the linear regression model also solves the same optimization problem, with a subtle, but important, difference in the constraint $\sum_j|\beta_j|\leq t$. It can be seen that no such constraints are imposed on the LSR. The RR uses a $l_2$-norm constraint on the $\boldsymbol{\beta}$, while LASSO employs the $l_1$-norm. Compared to RR, which could only shrink the coefficients, LASSO is able to zero out coefficients when $t$ is sufficiently small. Note that since the ridge regression employs an $l_2$-norm penalty, it thus has a closed form for the solution. Although LASSO does not have a closed form for its solution, there are several programs available for solving LASSO optimization, and here we employed the Glmnet for Matlab \cite{Friedman07,Friedman10}.

While LSR, RR and LASSO are point estimate algorithms, it is interesting to examine the BLASSO approach of \cite{Park08} which infers the posterior densities for the parameters $\boldsymbol{\beta}$. The detailed derivations for the conditionals can be found in \cite{Park08}, and the detailed Gibbs sampler can be found in the supplementary material.

\subsection*{Modelling negative regulations}
If $\boldsymbol{\beta}$ is non-negative, the negative interactions of miRNA and mRNA are modelled by the following linear regression model,
\begin{equation}
\mathbf{y} = -\mathbf{X}\boldsymbol{\beta} + \boldsymbol{\epsilon}
\end{equation}
where $\mathbf{X}$ is the miRNA expression data. To solve this model, we employ both a point estimate algorithm, i.e., the nLASSO, and the Bayesian solution, i.e., the nBLASSO, which will be described below.\\

The nLASSO for the linear regression model shares the same optimization problem with LSR, RR and LASSO, but with different constraints
$\beta_j\geq 0$ and $\sum_j\beta_j\leq t$.
One difference between LASSO and nLASSO is that nLASSO employs non-negative constraints. To implement this algorithm, we employ the Matlab software available at http://www.stanford.edu/$\sim$boyd/l1\_ls/ \cite{Kim07}.

\subsubsection*{Non-negative Bayesian LASSO}
Similar to the BLASSO, negative regulators can also be simulated by using sampling approaches. To simulate the required posterior distribution, the following data likelihood and priors are employed.
\begin{equation}
p(\mathbf{y}|\mathbf{X},\boldsymbol{\beta},\sigma^2)=\mathcal{N}(-\mathbf{X}\boldsymbol{\beta} , \sigma^2\mathbf{I})
\end{equation}
\begin{equation}
p(\boldsymbol{\beta}|\sigma^2,\lambda)=\prod_{m=1}^M\frac{\lambda_m}{2{\sigma^2}} \exp\left\{-\frac{\lambda_m \beta_m}{2{\sigma^2}}\right\}\mathbf{I}(\beta_m\geq 0)
\end{equation}
\begin{equation}
p(\lambda_m)=\frac{1}{\Gamma(\alpha_{\lambda}^0)(\beta^0_{\lambda})^{\alpha^0_{\lambda}}}\lambda_m^{\alpha^0_{\lambda}-1}e^{-\frac{\lambda_m}{\beta^0_{\lambda}}}\mathbf{I}(\lambda_m > 0)
\end{equation}
\begin{equation}
p(\sigma^2)\propto \frac{1}{\sigma^2}
\end{equation}

We also need to simulate the conditional distribution:
\begin{equation}
p(\boldsymbol{\beta},\sigma^2,\lambda|\mathbf{y},\mathbf{X})\propto p(\mathbf{y}|\mathbf{X},\boldsymbol{\beta},\sigma^2)p(\boldsymbol{\beta}|\sigma^2,\lambda)p(\sigma^2)p(\lambda)
\end{equation}
The conditional distribution for $\boldsymbol{\beta}$ is then a truncated normal distribution,
\begin{equation}
p(\boldsymbol{\beta}|\mathbf{y},\mathbf{X})\propto \mathcal{N}_+(\boldsymbol{\mu_{\beta}},\boldsymbol{\Sigma_{\beta}}), \beta_i\geq 0,
\end{equation}
where $\boldsymbol{\mu_{\beta}}=-\boldsymbol{{\Sigma}_{X}}(\mathbf{X}^T{\mathbf{y}}+\frac{1}{2}\boldsymbol{\lambda})$ and $\boldsymbol{\Sigma_{\beta}}=\sigma^2\boldsymbol{\Sigma_X}$ with $\boldsymbol{\Sigma_X}=(\mathbf{X}^T\mathbf{X})^{-1}$ and $\boldsymbol{\lambda}=(\lambda_1,\lambda_2,\cdots,\lambda_M)^T$. Sampling from the truncated normal distribution is shown in the following subsection.
The conditional distribution for $\sigma^{-2}\sim Gamma(\alpha_{\sigma},\beta_{\sigma})$ is a Gamma distribution where $\beta_{\sigma}=\left[\frac{1}{2}(\widetilde{\mathbf{y}}+\mathbf{X\boldsymbol{\beta}})^T(\widetilde{\mathbf{y}}+\mathbf{X\boldsymbol{\beta}})+\frac{1}{2}\sum_{m=1}^M\lambda_m\beta_m\right]^{-1}$ and $\alpha_{\sigma} = \frac{1}{2}N+M+2$.
The conditional for $\lambda_m$ is also a Gamma distribution, i.e., $\lambda_m\sim Gamma(\alpha_{\lambda},\beta_{\lambda})$, where $\alpha_{\lambda}=\alpha^0_{\lambda}+1$ and $\beta_{\lambda}=\left(\frac{1}{\beta^0_{\lambda}}+\frac{1}{2\sigma^2}\beta_m\right)^{-1}$.
So, finally, the Gibbs sampler in Algorithm 1 is used to simulate those parameters.
\begin{description}
\item[Algorithm 1] Gibbs Sampler for Non-negative Bayesian LASSO.
\item[Require:] $\mathbf{y}$, $\alpha_{\lambda}^0=1e^{-6}$, $\beta_{\lambda}^0=1e^6$, NSamps.
   \begin{description}
   \item[for] t=1:NSamps.
              \begin{description}
              \item[]Draw sample $\beta\sim\mathcal{N}_+(\mu_{\beta},\Sigma_{\beta})$.
              \item[]Draw sample $\sigma^{-2}\sim Gamma(\alpha_{\sigma},\beta_{\sigma})$.
              \item[]Draw sample $\lambda_m\sim Gamma(\alpha_{\lambda},\beta_{\lambda})$ for $m=1,2,\cdots,M$.
              \end{description}
   \item[end for]
   \end{description}
\end{description}

\subsubsection*{Sampling from the truncated normal distribution}
A Gibbs sampler is used to simulate the truncated normal distribution. Consider sampling from the general truncated normal distribution of the form
\begin{equation}
p(\mathbf{z}|\boldsymbol{\mu}, \boldsymbol{\Sigma})\propto \exp\left(-\frac{1}{2}(\mathbf{z}-\boldsymbol{\mu})^T\boldsymbol{\Sigma}^{-1}(\mathbf{z}-\boldsymbol{\mu})\right)\mathbf{I}(\mathbf{z}\in C)
\end{equation}
where the set $C$ denotes the support for the random variables $\mathbf{z}$. We need to sample $z_{i}$ given $z_{-i}=(z_1,z_2,\cdots,z_{i-1},z_{i+1},\cdots,z_M)$. The conditional distribution is then
\begin{equation}
p(z_i|z_{-i})\propto\exp\left(-\frac{1}{2v_i^2}(z_i-u_i)^2\right)I(z_i\in (c_i,d_i))
\end{equation}
where $v_i^2=\omega_{ii}^{-1}$, $u_i=\mu_i+\frac{1}{\omega_{ii}}\sum_{j\neq i}(\mu_j-z_j)\omega_{i,j}$, and $\omega_{ij}$ is the $ij$th element of $\boldsymbol{\Sigma}^{-1}$. Note that $c_i$ and $d_i$ are the end points of the support of $z_i$. We set $\xi=\frac{z_i-u_i}{v_i}$, and then simulate $p(\xi)\propto\exp(-\xi^2/2)I(\xi\in(c_{\xi},d_{\xi}))$ where $c_{\xi}=\frac{c_i-u_i}{v_i}$ and $d_{\xi}=\frac{d_i-u_i}{v_i}$. We employ the Gibbs sampler in \cite{Damien01} to simulate $p(\xi)$. Then $\xi$ is simulated in turn according to the conditionals which are uniformly distributed:
\begin{equation}
p(Y|\xi)\sim U(0,\exp(-\xi^2/2))
\end{equation}
\begin{equation}
p(\xi|y)\sim U\left(\max \left(c_{\xi},-\sqrt{-2\log{y}}\right),\min\left(d_{\xi},\sqrt{-2\log{y}}\right)\right)
\end{equation}

\subsubsection*{{A variable selection scheme for (non-negative) Bayesian LASSO}}
\begin{figure}[!ht]
  \centering
  \includegraphics[width=90mm]{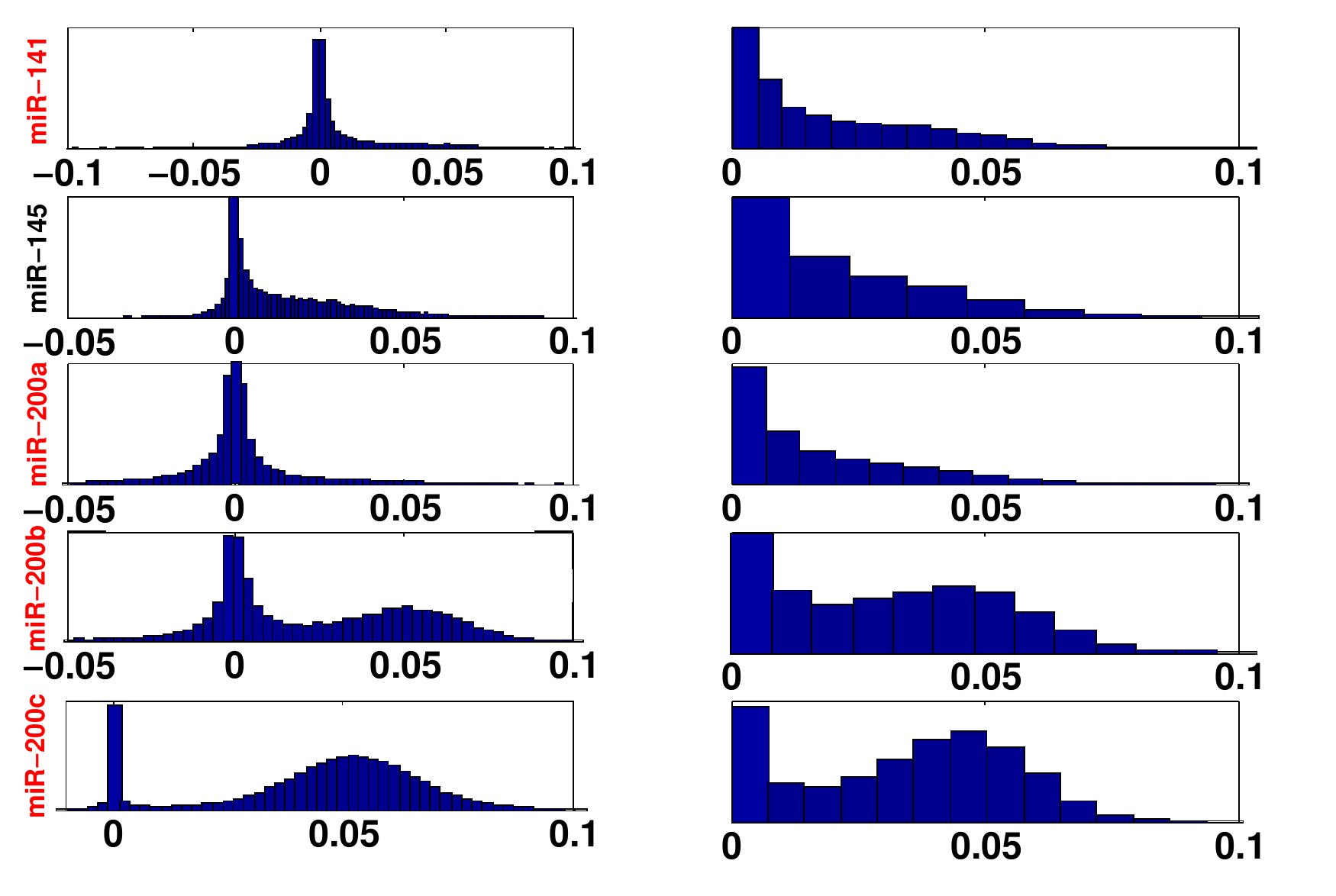}
  \caption{{\bf The inferred probability densities for $\boldsymbol{\beta}$ by using the Bayesian LASSO (left) and non-negative Bayesian LASSO (right).} By computing the active credible intervals defined for variable selections, nBLASSO identified these five miRNAs targeting the gene TWIST1 and BLASSO identified miR-145, 200b and 200c ($\alpha=0.05$). The miRNAs highlighted in red have been experimentally validated as down-regulating the gene TWIST1.}
  \label{bar_TWIST}
\end{figure}

Neither BLASSO nor nBLASSO are able to select variables automatically. A particular scheme for variable selection is thus proposed. In Bayesian statistics, a credible interval is usually used to summarize the statistics for the inferred posterior distribution of a parameter (see for example \cite{Eberly03}). Let us denote the credible interval for a variable by $[a, b]$. If $a$ is greater than zero, the variable is assumed to be active (or selected). We call it the active credible interval (ACI). Since many of the posterior distributions are multimodal, the posterior samples were partitioned into two clusters. For this particular problem, the posterior of $\beta_i$ has a high probability density around zero and can have another high density region away from zero (see Figures \ref{bar_TWIST}, \ref{bar_Notch_BLasso} \& \ref{bar_Notch_nBLasso}). Therefore, if the posterior has high probability density greater than zero, the corresponding variable is assumed to be active. Ideally, the samples of the posterior are partitioned into two clusters, where one cluster has a high probability density around zero and the other has such high densities elsewhere. For a variable $\beta_i$, suppose there are $T$ samples, denoted by $\beta_i^1,\beta_i^2,\cdots,\beta_i^T$. To calculate the ACI for $\beta_i$, the following procedure is performed. Firstly, we model the density for $\beta_i$ with a mixture distribution of the form $p(\beta_i)=\sum_{k=1}^2\pi_kf_k(\beta_i)$ where $f_k$ represents the component density and $\pi_k$ is the mixture weight. For the current problem, the $k$-means clustering algorithm is suitable for modelling this distribution and thus the samples were partitioned into two clusters. Given that the two clusters have means $m_1$ and $m_2$, suppose that $m_2>m_1$. Rank the samples for the cluster with mean $m_2$ and denote them by
$\beta_i^{(1)}\leq\beta_i^{(2)}\leq\cdots\leq\beta_i^{(t)}$. Secondly, calculate the $100(1-\tau)\%$ credible interval for the cluster $m_2$ and denote it by $[a,b]$. The estimate for $a$ is $\beta_i^{([t\frac{\tau}{2}]+1)}$ and the estimate for $b$ is $\beta_i^{([t(1-\frac{\tau}{2})]+1)}$, where $[x]$ denotes the largest integer less than or equal to $x$ and if $x$ is itself an integer, then $1$ is not added. Note that, in our experiments, $\tau=0.05$ was used to generate the credible intervals. If $a>0$, the API is thus $[a,b]$. The statistical significance of the ACI is then defined as
\begin{equation}
p=\frac{Q}{T}
\end{equation}
where $Q$ is the number of samples in the ACI and $T$ is the total number of the samples. For fixing a particular value $0<\alpha<1$, if $p>\alpha$, the variable is selected (or active). Obviously, the value of $\alpha$ controls the number of active variables. As the value of $\alpha$ increases, a smaller number of variables are selected.

\section*{Results}
In this section, six linear regression algorithms were applied to four expression datasets. The six linear regression algorithms are LSR, RR, LASSO, nLASSO, BLASSO and nBLASSO. The first four algorithms produce point estimates for the model and the last two are Bayesian methods.\\

\subsection*{Data and Experimental Settings}
\subsubsection*{Expression Datasets}
The experimentally validated targets of the four datasets were employed to evaluate the linear regression algorithms. The first, Multi Class Cancer (MCC), was introduced by \cite{Muniategui12a}. This dataset has 88 paired cancer and normal samples with mRNA data from \cite{Ram01} and miRNA expression data from \cite{Lu05}. We employed the method described in \cite{Muniategui12a} to select the genes. The initial set of interactions were generated from the union of miRBase, miRanda, miRGen, miRecords, TarBase and miRWalk. For this data, 23 mRNA--miRNA validated interactions found in the literature were used to validate these methods.

The following three datasets using Argonaute protein expressions were studied in \cite{Lu11} and \cite{Stanhope09}. The initial putative interactions were generated by the intersections of TargetScan and PicTar predictions. The second dataset we used here was originally used to study nasopharyngeal cancer (NPC) by researchers in Madison, WI, and was derived by profiling the mRNA and miRNA of 31 NPC and 10 normal tissule samples. The third dataset is from the Broad Institute and contains 67 tissue samples of 10 different normal and tumor tissue types. The fourth dataset is from a study conducted by the Memorial Sloan--Kettering Cancer Center (MSKCC) with 28 normal, 98 primary cancer and 13 metastatic cancer tissues samples using the Affymetrix Human Exon 1.0 ST Array and Agilent Human miRNA Microarray 2.0. We collected the experimentally identified mRNA--miRNA interactions by searching the literature for the datasets Broad, Madison and MSKCC, and also used the experimentally identified interactions collected by \cite{Lu11}. The numbers of the experimentally validated interactions are listed in Table \ref{final_res}. The experimentally validated interactions which were included in the putative interactions were used for assessing the algorithms. The described algorithms were applied to the initial set of putative interactions to infer the mRNA--miRNA interactions. It should be mentioned that the number of interactions should be smaller than the number of samples in order to use LSR. 

\subsubsection*{{Selection of the regulation parameter $\lambda$ and the coefficients $\boldsymbol{\beta}$}}
RR, LASSO and the nLASSO have a regulation parameter $\lambda>0$ (or $t>0$), which requires tuning. The penalty parameter $\lambda$ controls the amount of the shrinkage of $\boldsymbol{\beta}$. The parameter $\lambda$ was determined by testing the values $\lambda_j=\left(\frac{j}{j_0}\right)^c$ where $j_0$ was some real number, $c$ a small integer,  and $j$ is in the range $[1,aj_0]$ where $a$ is an integer.
The larger  the value of $\lambda$ (or the smaller the value of $t$), the greater the amount of the shrinkage.

In order to select the regulation parameter, we used $K$-fold Cross-Validation (CV) \cite{Hastie01}. $K$-fold CV separates the data into $K$ roughly-equally sized parts, and uses $K-1$ parts as the training data to make the optimization. To partition the data, we set $K=11$ for the MCC, Broad and MSKCC data, and $K=10$ for the Madison data.

After the model has been trained, the estimated parameters are used to calculate the test error for the part which was not used for training. We select $\lambda$ for each of the mRNA samples. Thus for the expression data $\mathbf{y}=(y_1,y_2,\cdots,y_N)^T$ of one gene, $\mathbf{y}$ was divided into $K$ parts. We denote the part which was not used for training by $\mathbf{y}_{test}$. Similar partitions for $\mathbf{X}$ were performed. The corresponding testing part was denoted by $\mathbf{X}_{test}$. The test error $\widehat{\epsilon}_{test}=\left(\mathbf{y}_{test}-\mathbf{X}_{test}\boldsymbol{\beta}^*\right)^T\left(\mathbf{y}_{test}-\mathbf{X}_{test}\boldsymbol{\beta}^*\right)$
was then computed, where $\boldsymbol{\beta}^*$ denotes an estimate for $\boldsymbol{\beta}$ using the training data. Finally, the optimized regulation parameter was selected as the one which minimizes the mean test error.

It should be emphasized that $\mathbf{y}$ represents the expression data for a particular gene and each column of $\mathbf{X}$ represents the expression data for one microRNA. The $\beta_i$ is the corresponding coefficient of $\mathbf{x}_i$. In our representation, a positive $\beta_i$ denotes down-regulation. So a larger value for $\beta_i$ implies that the corresponding miRNA  has more impact on the gene. A subset of the miRNA that down-regulates the gene needs to be identified. After the regulation parameters were selected, the coefficients $\boldsymbol{\beta}$ were estimated by fixing the selected regulation parameters. For the point estimate algorithms LSR, RR, LASSO and nLASSO, a threshold was fixed to select the miRNAs that down-regulate the gene. LASSO chose $\beta_i=0$ for most $i$, so the threshold for LASSO could be set to be zero. For the LSR, RR and nLASSO, the threshold had to be larger than zero. If a $\beta_i$ was larger than the threshold, it was selected, and the corresponding miRNA was marked as a possible gene down-regulator.

\begin{figure}[!ht]
  \centering
  \includegraphics[width=90mm]{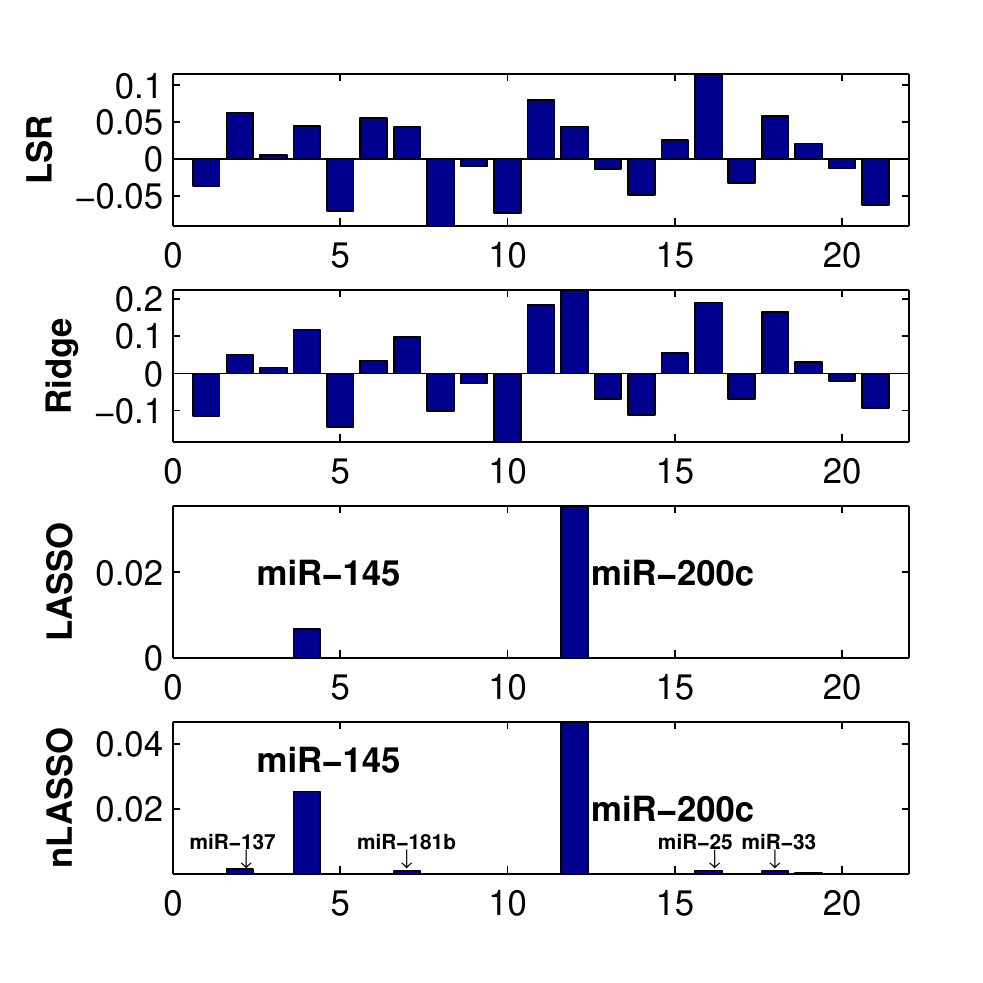}
  \caption{{\bf The point estimates for $\boldsymbol{\beta}$ for the gene TWIST1 in the MCC dataset using Least Squares Regression, Ridge Regression, LASSO and Non-negative LASSO algorithms.} The LSR and Ridge regression algorithms produced similar results for estimating mRNA and miRNA interactions, while LASSO and nLASSO produced sparse representations for the interactions.}
  \label{bars_lsr_ridge_lasso_nlasso}
\end{figure}

\begin{figure}[!ht]
  \centering
  \includegraphics[width=90mm]{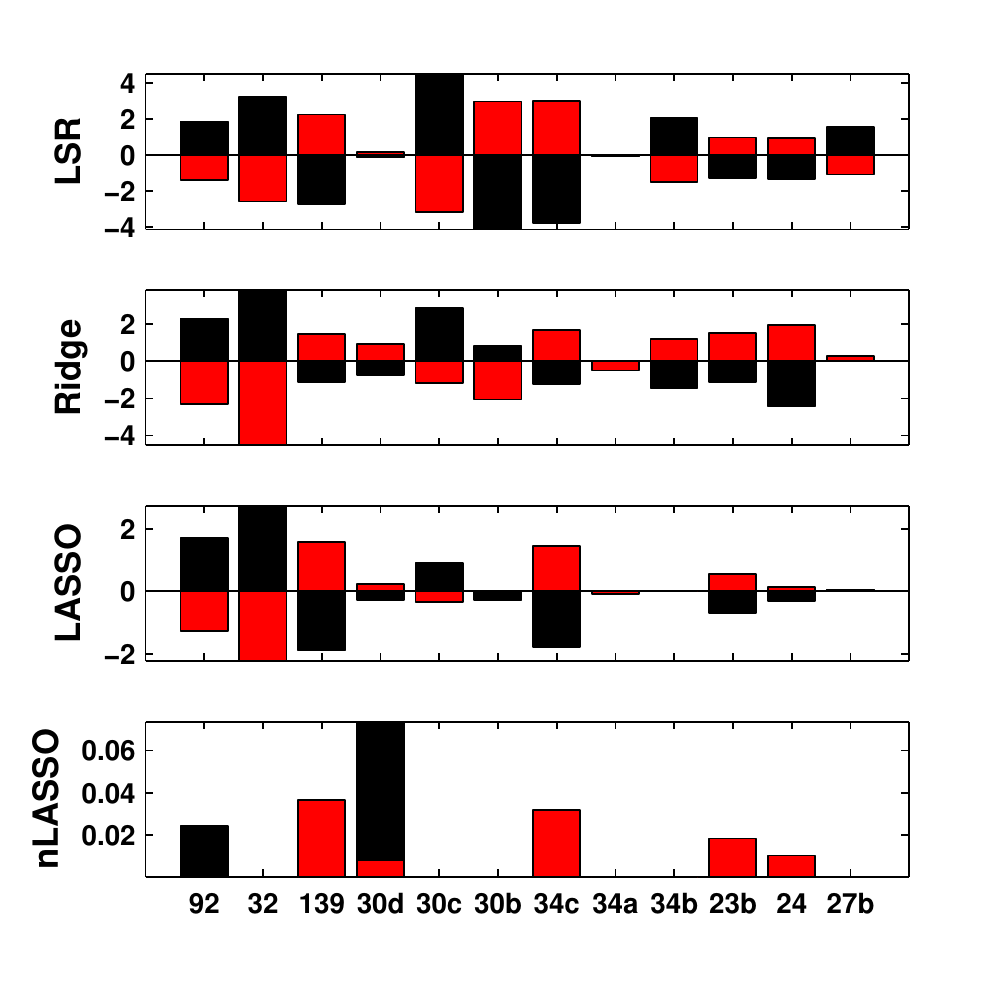}
  \caption{{\bf The point estimates for $\boldsymbol{\beta}$ using Least Squares Regression, Ridge Regression, the LASSO, and the Non-negative LASSO algorithms for the gene NOTCH1 using Argonaute expression}. Black bars denote the values for miRNA using Argonaute 2 and red for Argonaute 1, 3, and 4. The LSR and ridge regression algorithms produced similar results for $\boldsymbol{\beta}$. LASSO and nLASSO produced sparse representations. }
  \label{bar_argo_notch1}
\end{figure}

{\subsection*{The Methods with Non-negative Constraints Perform Better in Modelling Down-regulating Interactions for mRNA and miRNA}}
The six linear regression algorithms were applied to the four datasets described. 
The four algorithms, i.e. LSR, RR, LASSO and nLASSO, produce point estimates for those parameters. 
Across all datasets, LSR and RR, which produced similar results, were not able to learn sparse interactions. 
Both LASSO and nLASSO were able to learn sparse models. For most of the cases, 
LASSO shrunk most of the coefficients to zero, while all the coefficients estimated by nLASSO were, 
in our experiments, non-zero. See Figures \ref{bars_lsr_ridge_lasso_nlasso} \& \ref{bar_argo_notch1} for examples. 
The gene TWIST was used as an example to see how those algorithms were used for inferring the interactions. 
Figure \ref{bars_lsr_ridge_lasso_nlasso} shows the learned $\beta$ for the gene TWIST1 where there were 21 candidate miRNAs.
 Both LASSO and nLASSO identified miR-145 and 200c confidently. 
All of the other miRNAs except these two were estimated as exactly zero by LASSO. Several other targets, 
such as miR-137, were also detected by nLASSO, 
but were assigned less importance (see Figure \ref{bars_lsr_ridge_lasso_nlasso}). 
On the other hand, LSR and RR both identified 11 interactions. 
Therefore, these results show that both LASSO and nLASSO are able to learn sparse interactions. \\

Both LASSO and nLASSO confidently identified miR-145 and 200c for TWIST1; however, since nLASSO also assigned a small confidence value to several other miRNAs, a threshold had to be set for accurate selection. Unlike these point estimate algorithms, BLASSO and nBLASSO enable us to observe the posterior densities for those parameters for the gene TWIST1. These estimated densities were then used to compute the active credible intervals for those coefficients. The statistical significance for the active credible intervals, which guided us in selecting the interactions, was subsequently calculated. This procedure measures the uncertainty for the inferred interactions, 
for example, Figure \ref{bar_TWIST} shows the inferred densities for five miRNAs. By looking at the densities inferred by BLASSO, we see that it was confident in selecting miR-145, miR-200b and miR-200c and ignoring miR-141 and miR-200a since the densities of the latter are symmetric around zero. The densities inferred by nBLASSO show that we would confidently select miR-200b and miR-200c since their densities have two peaks, one around zero and the other around 0.05, and that it is uncertain about the significance of miR-141, miR-145 and miR-200a since they have only one peak at around zero. We are then required to calculate the active credible intervals and their statistical significance to perform the variable selections. For the gene TWIST1, the results for the active credible intervals and their statistical significance are shown in Table \ref{table_BLasso_TWIST} for BLASSO and Table \ref{table_nBLasso_TWIST} for nBLASSO. If $\alpha=0.05$, BLASSO selected miR-145, miR-200b and miR-200c; nBLASSO selected miR-141, miR-145, miR-200a, miR-200b and miR-200c. It should be noted that the miR-200 family (miR-200a, miR-200b, miR-200c and miR-141) have been experimentally identified as down-regulating TWIST1 \cite{Wik11}. nBLASSO is shown to be able to identify all the validated interactions for the current example.


All the genes were analysed using the same procedure. We show all the credible intervals for which the statistical significance was greater than $\alpha=0.05$ for the genes of the MCC dataset, as computed by BLASSO and nBLASSO in Tables \ref{table_mcc1}
\& \ref{table_nblasso_mcc}, respectively. 
Note that to perform the variable selection, a threshold has to be chosen for the point estimate algorithms. It can be seen that nBLASSO selected more miRNAs than BLASSO, which may be due to the non-negative constraints. We also counted the number of experimentally validated interactions estimated by the LASSO, nLASSO, BLASSO and nBLASSO for all four datasets, which are shown in Table \ref{final_res}. The threshold was set to be $0$ for LASSO and various values were set for nLASSO. Table \ref{final_res} demonstrates that it is difficult to find a routine for setting a reasonable threshold for nLASSO. For the Bayesian methods, we set $\alpha=0.05$ as the threshold for statistical significance. For the BLASSO and nBLASSO, it is natural and meaningful to use a statistical significance as the threshold, which provides the uncertainty of the predictions.\\

\begin{figure}[!ht]
  \centering
  \includegraphics[width=90mm]{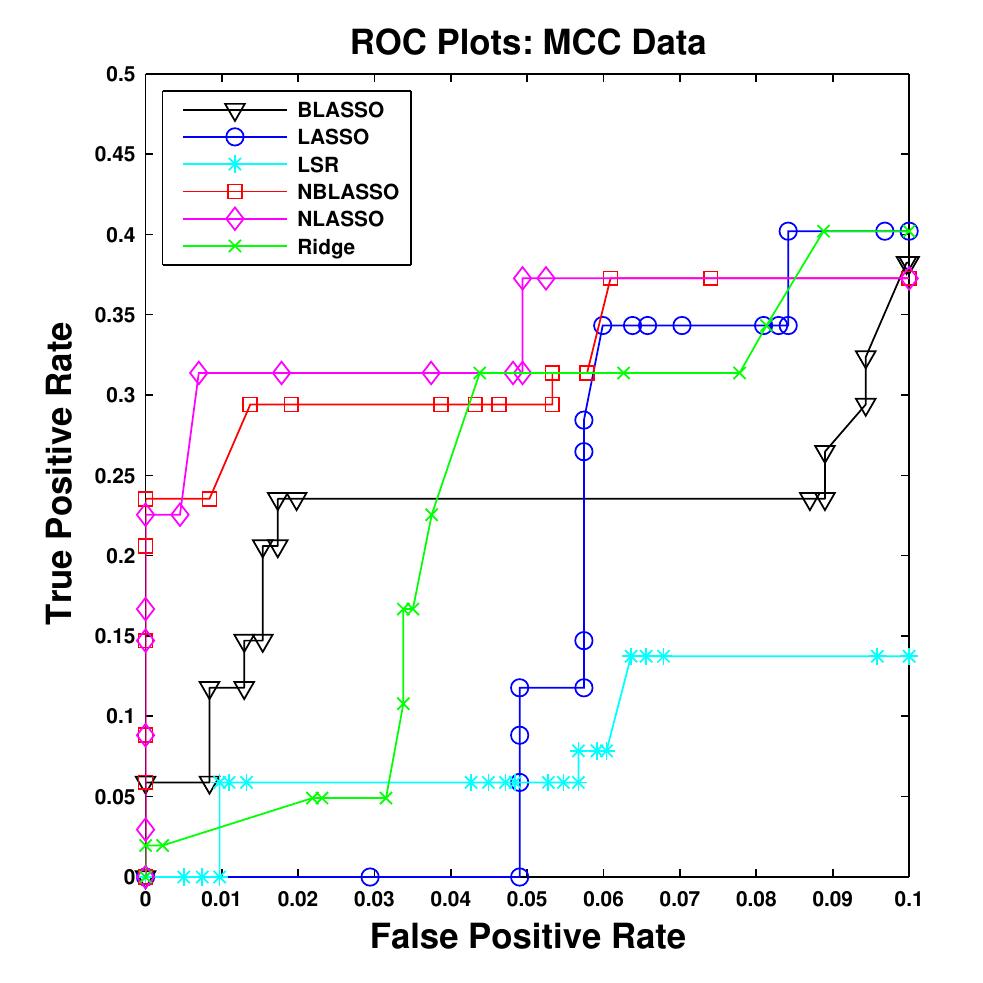}
  \caption{{\bf ROC plots for the MCC data}. The ROC values were produced for BLASSO and nBLASSO when fixing various values for the $\alpha=0.01,0.02,\cdots,0.9$. The ROC values for the LSR, RR, LASSO and nLASSO were computed when various thresholds were fixed. The threshold was set to $0, 1e-8, 1e-7, \cdots, 1e-3, 2e-3, 3e-3,\cdots, 1e-2, \cdots$. }
  \label{roc_mcc}
\end{figure}

\begin{figure}[!ht]
  \centering
  \includegraphics[width=90mm]{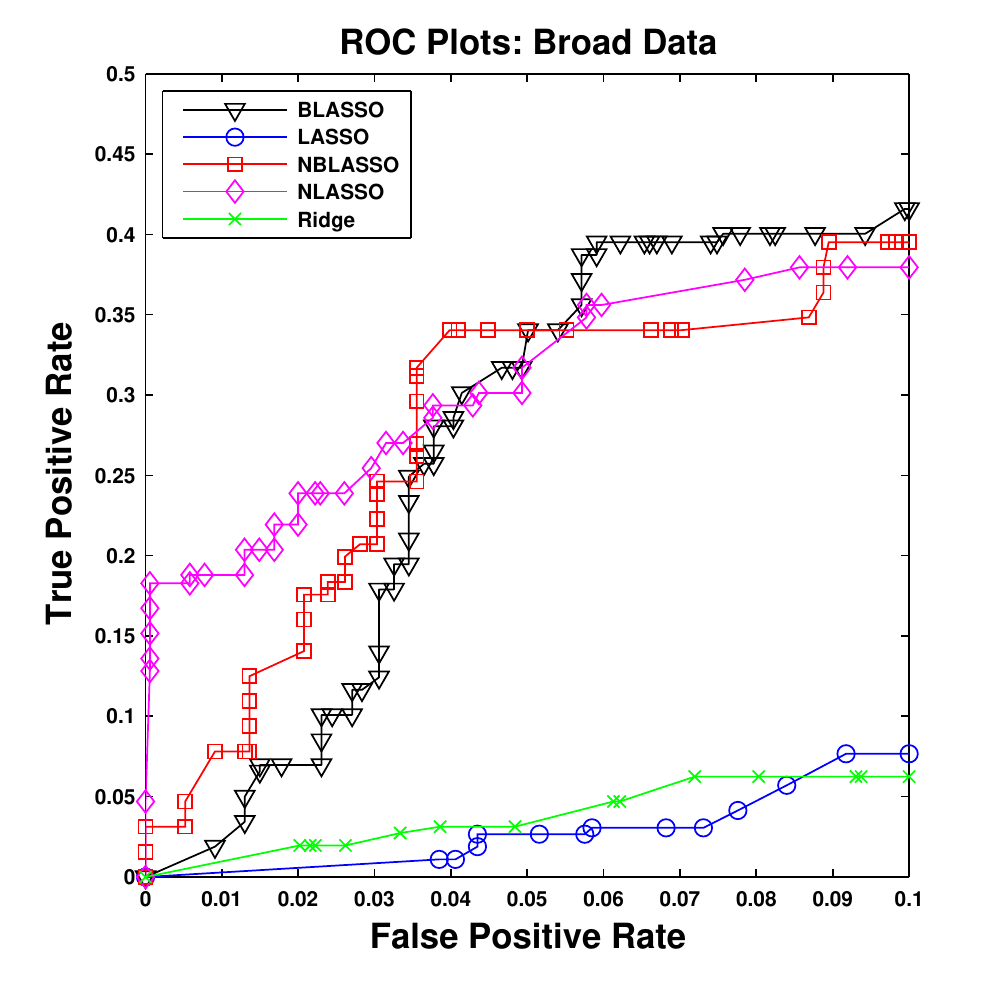}
  \caption{{\bf ROC plots for the Broad data}.}
  \label{roc_broad}
\end{figure}

\begin{figure}[!ht]
  \centering
  \includegraphics[width=90mm]{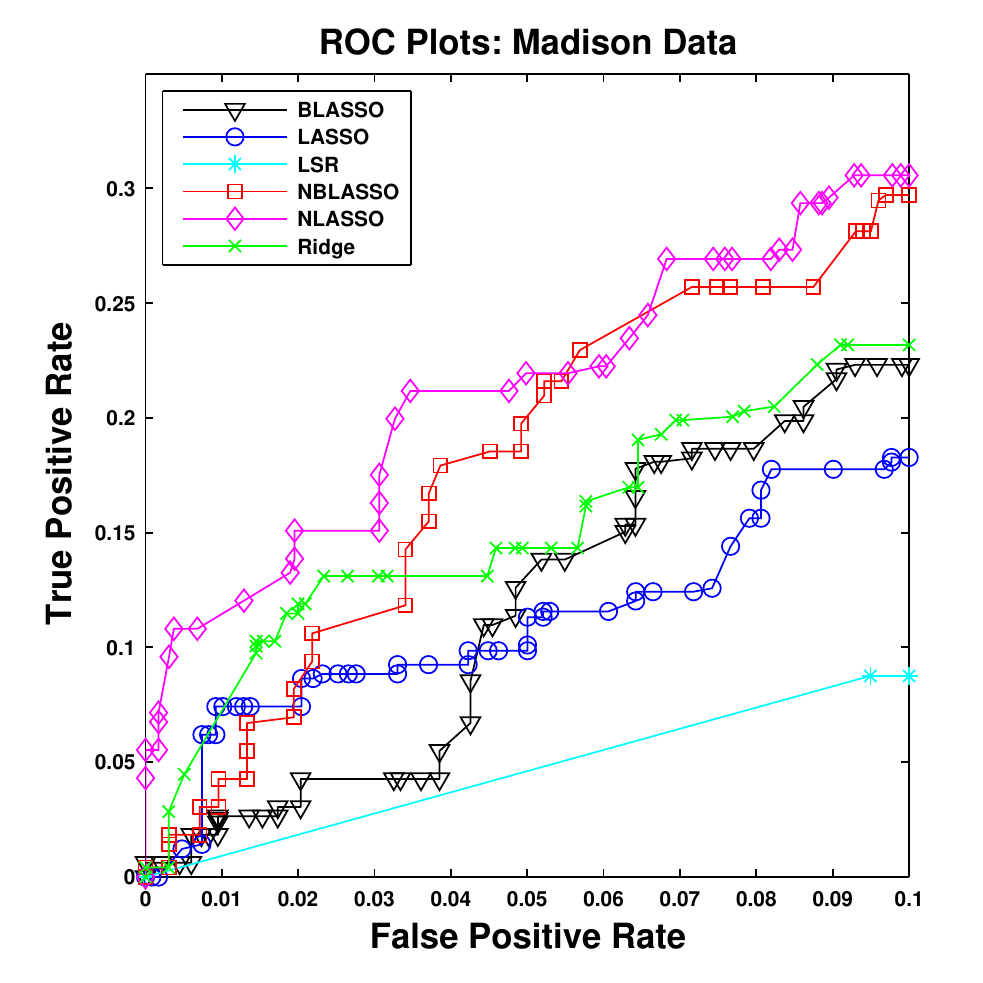}
  \caption{{\bf ROC plots for the Madison data}.}
  \label{roc_madison}
\end{figure}

\begin{figure}[!ht]
  \centering
  \includegraphics[width=90mm]{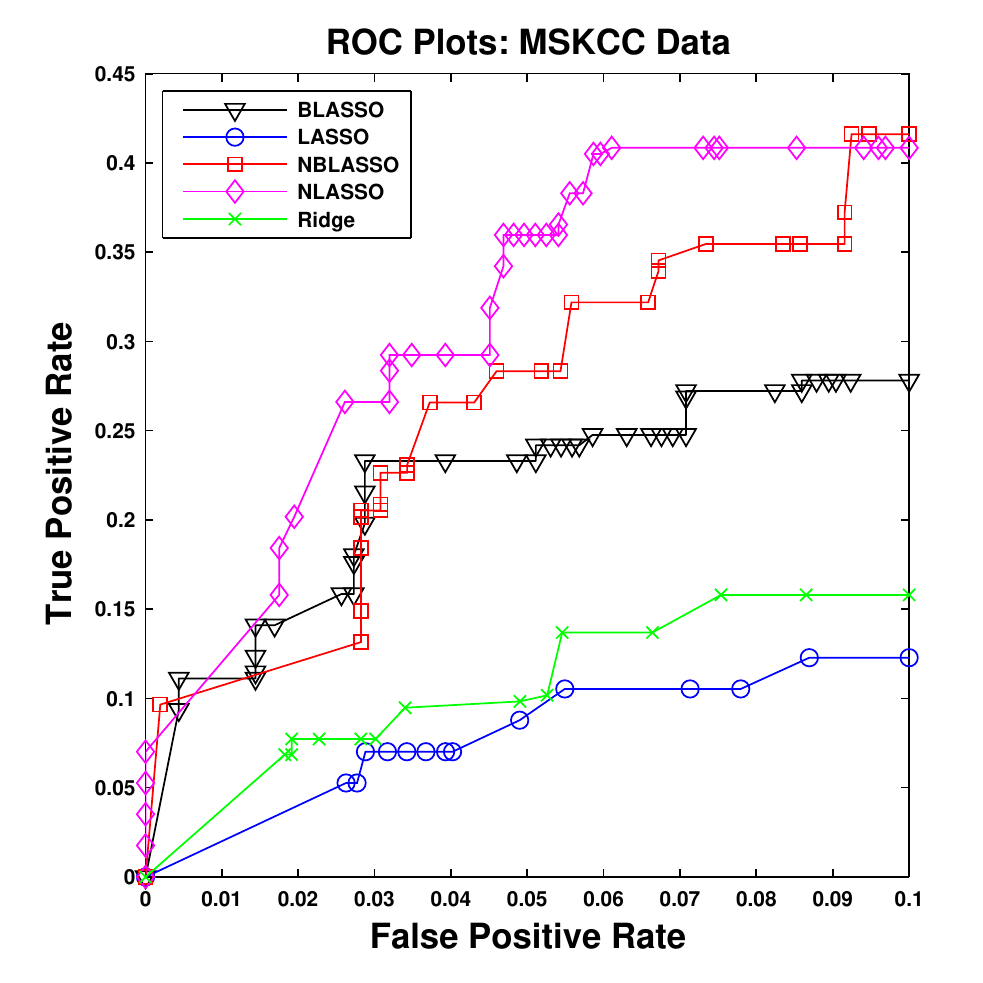}
  \caption{{\bf ROC plots for the MSKCC data}.}
  \label{roc_mskcc}
\end{figure}

The performance of these algorithms was compared by computing their sensitivities and specificities. To produce the ROC curves, various values for $\alpha$ were chosen for the Bayesian methods. The $\alpha$ could be varied from zero to one by a step size such as $0.01$. For the point estimate algorithms, a threshold is also required. The threshold was set to be $0, 1e^{-8}, 1e^{-7},\cdots,1e^{-3},2e^{-3},3e^{-3},\cdots,1e^{-2},2e^{-2},3e^{-2},\cdots$. Given these values, the ROC curves for all the datasets are plotted in Figures \ref{roc_mcc}, \ref{roc_broad}, \ref{roc_madison} and \ref{roc_mskcc}. Only the ROC of the false positive range $[0,0.1]$ are shown, for the purposes of comparison. For all four datasets, nBLASSO and nLASSO perform best in terms of both sensitivity and specificity. The performance of BLASSO was close to that of nBLASSO. The reason that LASSO was not better than nLASSO might be that LASSO set most of the coefficients to be zero,  while the coefficients estimated by nLASSO were all non-zero. As was noted, another reason might be that nLASSO uses non-negative constraints. The areas under the ROC curves (AUC) in the false positive range $[0,0.1]$ were also computed, and the results are shown in Table \ref{auc}. For the point estimate algorithms, the AUC values of nLASSO were larger than those of LSR, RR and LASSO across all datasets. For the Bayesian methods, the AUC values of nBLASSO were larger than those of BLASSO across all the data. The nLASSO performed better than nBLASSO in terms of the AUC. However, it has been noted that a meaningful criterion needs to be defined for nLASSO to select the threshold $\beta$. Overall, in terms of the AUC, nBLASSO and nLASSO performed better than the methods without non-negative constraints.\\

\subsection*{{{Inferring mRNA--miRNA Interactions Using Argonaute Protein Information}}}
It has been shown that the regression models employing Argonaute expressions fit the expression profiles of known targets pairs substantially better than models based only on miRNA expression data \cite{Stanhope09,Lu11}. All six algorithms were applied to three expression datasets, i.e., Broad, Madison and MSKCC, involving Argonaute expression information, which were also studied in \cite{Stanhope09} and \cite{Lu11}.

Here we discuss the results for the gene NOTCH1 in detail, although all other genes were analysed using the same procedure. Twelve miRNAs were selected as the candidates for interacting with the gene NOTCH1. Six miRNAs have been experimentally detected to down-regulate NOTCH1: miRNA 34a, 34b \& 34c, miRNA 23b, miRNA 24 and miRNA 27b \cite{Fukuda05,Wang10,zheng12,Melo12}. The point estimate algorithms, i.e., the LSR, RR, LASSO and nLASSO, were applied to the expression data to identify the interactions of the mRNA and miRNA. The estimated $\boldsymbol{\beta}$ are shown in Figure \ref{bar_argo_notch1}. In the figure, the black bars plot the $\boldsymbol{\beta}$ for the miRNA involving Argonaute 2 expression information and the red bars plot the $\boldsymbol{\beta}$ for the miRNA involving Argonaute 1,3\&4 expression information. Obviously the LSR and RR produced similar results. Since the Argonaute 2 protein is the competitor of Argonaute 1,3\&4 in forming the RISC for targeting mRNA, as was discussed in \cite{Stanhope09}, the values of $\boldsymbol{\beta}$ for miRNA with Ago2 information computed by LSR, RR and LASSO are inversely related to those of miRNA with Ago1,3,\&4 information in sign with similar magnitudes of the $\boldsymbol{\beta}$ values. For the results of the nLASSO, if the effect of an miRNA with Ago2 is positive, the effect of the miRNA with Ago1,3\&4 then is around zero, and vice versa, unless they are both zero. For the LSR and RR, we think it would be better to select all of the candidate miRNAs since all of the inferred coefficients were non-zero. As expected, both LASSO and nLASSO select smaller numbers of miRNAs.

\begin{figure}[!ht]
  \centering
  \includegraphics[width=90mm]{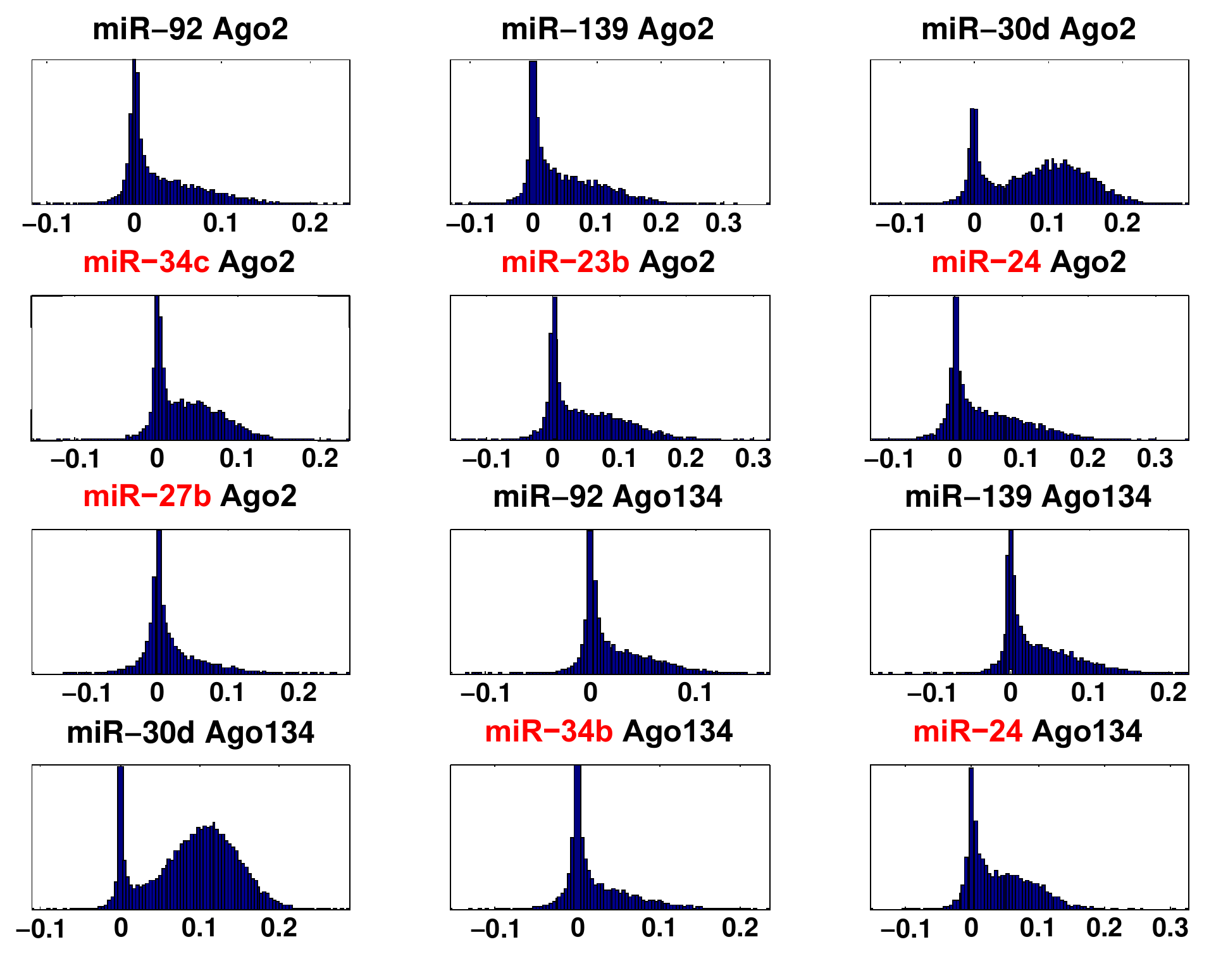}
  \caption{{\bf The probability densities for $\boldsymbol{\beta}$ inferred by using the Bayesian LASSO for the gene NOTCH1}. All the listed miRNAs were detected as targeting the gene NOTCH1 by computing the active credible intervals proposed for variable selection ($\alpha=0.05$). MiRNA-23b, miRNA-24, miRNA-27b, miRNA-34b and miRNA-34c have been experimentally validated as down-regulating the gene NOTCH1. }
  \label{bar_Notch_BLasso}
\end{figure}

\begin{figure}[!ht]
  \centering
  \includegraphics[width=90mm]{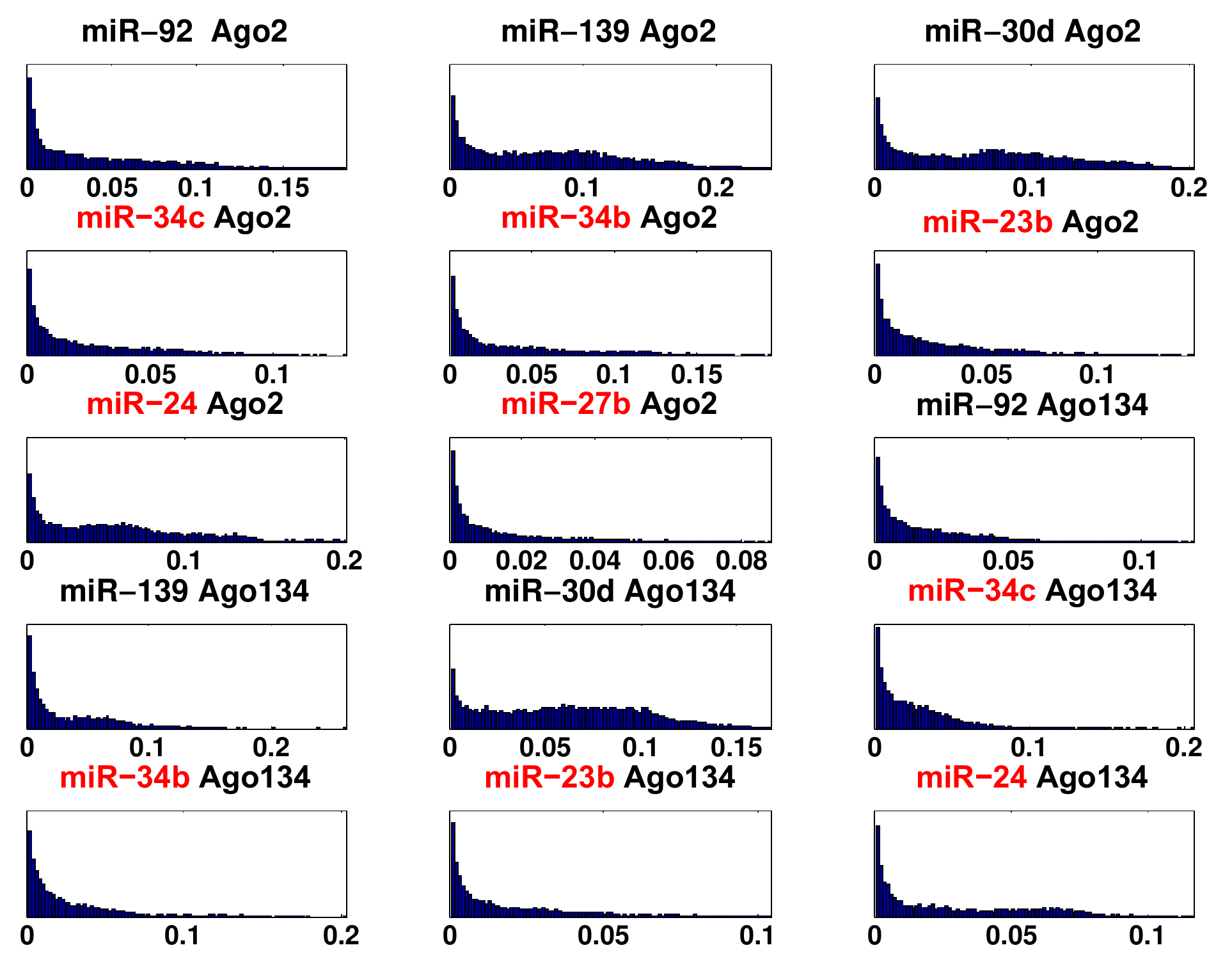}
  \caption{{\bf The probability densities for $\boldsymbol{\beta}$ inferred by using the non-negative Bayesian LASSO for the gene NOTCH1}. All the listed miRNAs were detected targeting the gene NOTCH1 by computing the active credible intervals proposed for variable selection ($\alpha=0.05$). MiRNA-23b, miRNA-24, miRNA-27b, miRNA-34b and miRNA-34c have been experimentally validated as down-regulating the gene NOTCH1.}
  \label{bar_Notch_nBLasso}
\end{figure}

\begin{figure}[!ht]
  \centering
  \includegraphics[width=90mm]{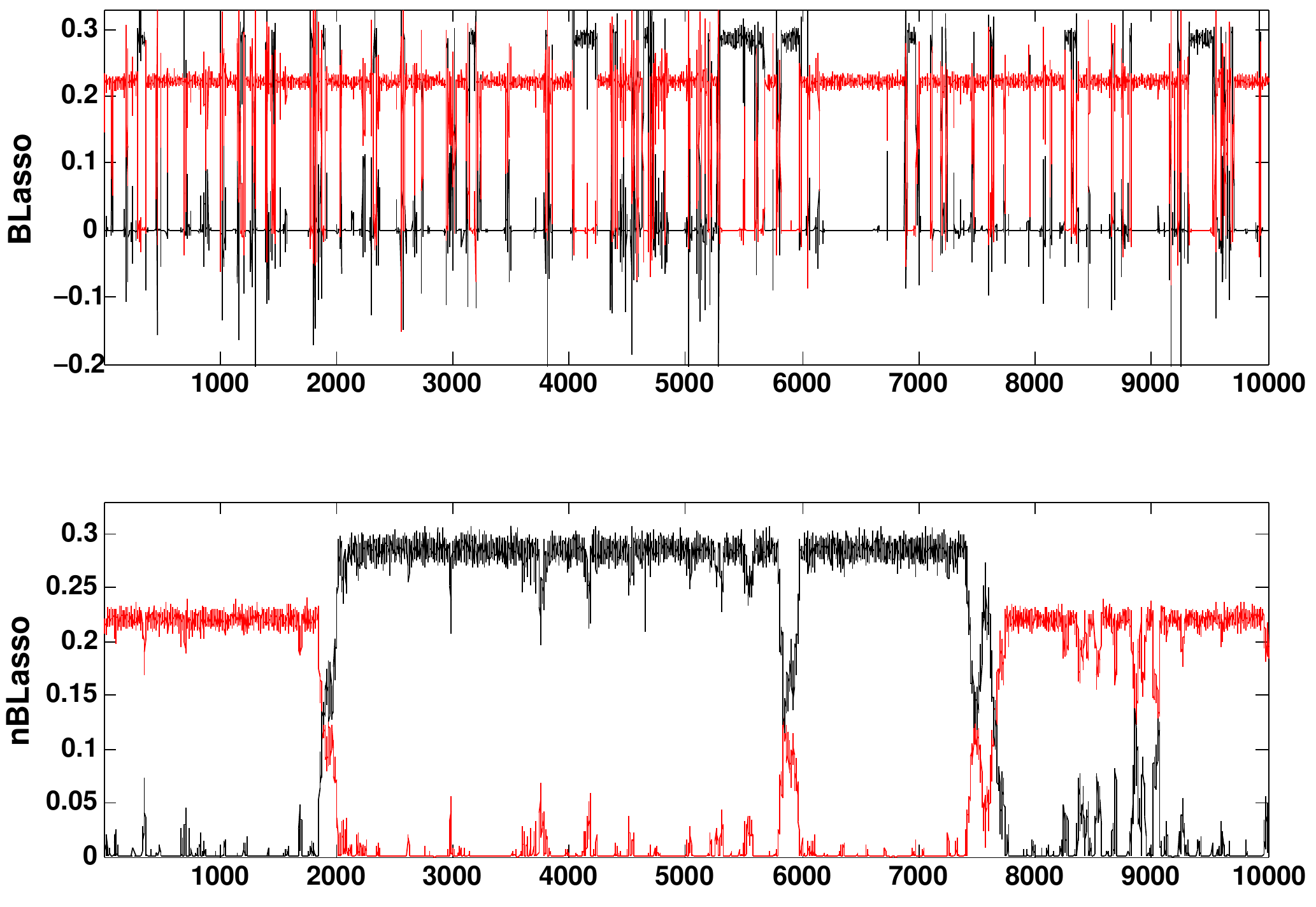}
  \caption{{\bf The samples for $\boldsymbol{\beta}$ drawn by using Bayesian LASSO and non-negative Bayesian LASSO for the gene ACAA2}. Only the miRNA-124 was considered as the candidate for targeting ACAA2. The RISCs miRNA-124$\times$Ago2 and miRNA-124$\times$Ago134, which are presumably competing, were simultaneously put into the model for interacting with ACAA2. The black plots the $\beta$ for miRNA-124$\times$Ago2 and the red plots for miRNA-124$\times$Ago134, showing that when one of the RISCs is on for targeting ACAA2, the other is off. }
  \label{B_trace_ACAA2}
\end{figure}

The Bayesian approaches were then applied to this data. The probability densities are plotted in Figures \ref{bar_Notch_BLasso} \& \ref{bar_Notch_nBLasso} for BLASSO and nBLASSO, respectively. The calculated active credible intervals and their statistical significance are shown in Tables \ref{table_bLasso_Notch} \& \ref{table_nbLasso_Notch}. The parameter $\alpha=0.05$ was used. BLASSO selected seven miRNAs combined with Ago2 and five miRNAs combined with Ago1,3\&4, which contains 5 known targets out of 6. nBLASSO selected eight miRNAs combined with Ago2 and seven miRNAs combined with Ago1,3\&4, which also contains 5 out of 6 known targets. Their densities are shown in Figures \ref{bar_Notch_BLasso} \& \ref{bar_Notch_nBLasso}. For the purpose of clearly comparing the effects of Ago2 and Ago1,3\&4, the results for gene ACAA2 are shown, where only miRNA-124 was selected as the candidate for targeting ACAA2. Figure \ref{B_trace_ACAA2} plots the traces for sampling the effects of miRNA-124$\times$Ago2 and miRNA-124$\times$Ago1,3\&4. Interestingly, when the effects of miRNA-124$\times$Ago2 are on, the effects of miRNA-124$\times$Ago1,3\&4 are then off and vice versa. This implies that Ago2 and Ago1,3\&4 compete with each other to form the RISC, which agrees with the results of \cite{Stanhope09}.



\section*{Conclusions}
Linear regression models have been proved suitable for modelling miRNA targets. In this paper, we have proposed BLASSO and nBLASSO to infer mRNA--miRNA interactions. LSR, RR, LASSO, nLASSO, BLASSO and nBLASSO have been applied to four publicly available expression datasets. We concluded that the LSR and RR algorithms have similar performances. LASSO and nLASSO produce sparse representations for the mRNA--miRNA interactions. It has been demonstrated that nLASSO and nBLASSO perform best in terms of the AUC. However, it should be noted that LSR and RR estimate more mRNA--miRNA interactions than LASSO and nLASSO. Both the BLASSO and nBLASSO provide uncertainties for the estimates of the interactions. The Bayesian methods do not perform variable selections automatically and thus the credible intervals were employed to accomplish this task by looking at a statistical significance. Of course, the point estimate algorithms also lack automatic variable selection and thus a threshold had to be manually fixed for selecting the miRNAs. However, advanced methodologies could be employed for performing variable selections for this problem, as was proposed in \cite{zhong11}. We prefer the Bayesian approaches rather than the point estimate algorithms, because Bayesian methods produce the probability densities for the effects of the miRNAs on their targets. Accordingly, the mRNA--miRNA interactions can be analysed by graphing those inferred densities of the effects. It has been well known that the Ago2 proteins are competitors of the Ago1,3\&4 proteins when binding with RISC. Our Gibbs sampler was able to simulate this competition by sampling the effects of the RISC on the miRNA targets.

\section*{Acknowledgements}
We would like to thank Tamara Polajnar (University of Cambridge, Computer Laboratory)
and David Sterratt (School of Informatics, University of Edinburgh) 
for their help in the final stages of the preparation of this manuscript.



\bibliographystyle{plain}
\bibliography{RJ-Lasso}


\newpage

\begin{table}[!ht]
\caption{\bf{Statistical summary for estimated effects of miRNAs by Bayesian LASSO.}}
\begin{tabular}{|c|c|c|c|}
  \hline
  mRNA & miRNA & Significance & Credible Interval \\
  \hline
  TWIST1 & miR-145 & 0.104 & [0.018,0.068] \\
  \cline{2-4}
   & \textcolor[rgb]{1.00,0.00,0.00}{miR-200b} & 0.153 & [0.028,0.083] \\
  \cline{2-4}
   & \textcolor[rgb]{1.00,0.00,0.00}{miR-200c} & 0.691 & [0.032,0.086] \\
  \hline
\end{tabular}
\begin{flushleft}The active credible intervals and their statistical significance computed by Bayesian LASSO for gene TWIST1. Three miRNAs were identified by Bayesian LASSO as targeting TWIST1. Only the miRNAs with significance greater than $0.05$ are shown. The miRNAs highlighted in red have been experimentally shown to target TWIST1.
\end{flushleft}
\label{table_BLasso_TWIST}
\end{table}

\vspace{-0.5cm}
\begin{table}[!ht]
\caption{\bf{Statistical summary for estimated effects of miRNAs by non-negative Bayesian LASSO.}}
\begin{tabular}{|c|c|c|c|}
  \hline
  mRNA & miRNA & Significance & Credible Interval \\
  \hline
  TWIST1 & \textcolor[rgb]{1.00,0.00,0.00}{miR-141} & 0.073 & [0.023,0.113] \\
  \cline{2-4}
   & miR-145 & 0.104 & [0.026,0.132] \\
  \cline{2-4}
   & \textcolor[rgb]{1.00,0.00,0.00}{miR-200a} & 0.058 & [0.023,0.118] \\
  \cline{2-4}
   & \textcolor[rgb]{1.00,0.00,0.00}{miR-200b} & 0.193 & [0.028,0.111] \\
   \cline{2-4}
   & \textcolor[rgb]{1.00,0.00,0.00}{miR-200c} & 0.390 & [0.028,0.096] \\
  \hline
\end{tabular}
\begin{flushleft}The active credible intervals and their statistical significance computed by non-negative Bayesian LASSO for gene TWIST1. Six miRNAs were identified by non-negative Bayesian LASSO as targeting TWIST1. Only the miRNAs with significance greater than $0.05$ are shown. The miRNAs highlighted in red have been experimentally shown to target TWIST1.
\end{flushleft}
\label{table_nBLasso_TWIST}
\end{table}

\begin{table}[!ht]
\caption{\bf{Statistical summary for estimated effects of RISCs by Bayesian LASSO.}}
\begin{tabular}{|c|c|c|c|}
  \hline
  mRNA & RISC & Significance & Credible Interval \\
  \hline
  NOTCH1 & miR-92$\times$Ago2 & 0.058 & [0.048,0.217] \\
  \cline{2-4}
   & miR-139$\times$Ago2 & 0.056 & [0.054,0.171] \\
  \cline{2-4}
  & miR-30d$\times$Ago2 & 0.162 & [0.071,0.198] \\
  \cline{2-4}
  & \textcolor[rgb]{1.00,0.00,0.00}{miR-34c}$\times$Ago2 & 0.262 & [0.041,0.138] \\
  \cline{2-4}
  & \textcolor[rgb]{1.00,0.00,0.00}{miR-23b}$\times$Ago2 & 0.182 & [0.064,0.203] \\
  \cline{2-4}
  & \textcolor[rgb]{1.00,0.00,0.00}{miR-24}$\times$Ago2 & 0.234 & [0.067,0.220] \\
  \cline{2-4}
  & \textcolor[rgb]{1.00,0.00,0.00}{miR-27b}$\times$Ago2 & 0.066 & [0.039,0.154] \\
  \cline{2-4}
  & miR-92$\times$Ago134 & 0.078 & [0.024,0.086] \\
  \cline{2-4}
  & miR-139$\times$Ago134 & 0.154 & [0.046,0.167] \\
  \cline{2-4}
  & miR-30d$\times$Ago134 & 0.530 & [0.071,0.191] \\
  \cline{2-4}
  & \textcolor[rgb]{1.00,0.00,0.00}{miR-34b}$\times$Ago134 & 0.118 & [0.047,0.146] \\
  \cline{2-4}
  & \textcolor[rgb]{1.00,0.00,0.00}{miR-24}$\times$Ago134 & 0.138 & [0.051,0.207] \\
  \hline
\end{tabular}
\begin{flushleft}The active credible intervals and their statistical significance computed by Bayesian LASSO for gene NOTCH1. Seven miRNAs were identified by Bayesian LASSO as targeting NOTCH1. Only the miRNAs with significance greater than $0.05$ are shown.  The miRNAs highlighted in red have been experimentally validated as down-regulating NOTCH1.
\end{flushleft}
\label{table_bLasso_Notch}
\end{table}

\begin{table}[!ht]
\caption{\bf{Statistical summary for estimated effects of RISCs by non-negative Bayesian LASSO.}}
\begin{tabular}{|c|c|c|c|}
  \hline
  mRNA & RISC & Significance & Credible Interval \\
  \hline
  NOTCH1 & miR-92$\times$Ago2 & 0.124 & [0.046,0.156] \\
  \cline{2-4}
   & miR-139$\times$Ago2 & 0.175 & [0.061,0.203] \\
  \cline{2-4}
  & miR-30d$\times$Ago2 & 0.249 & [0.059,0.177] \\
  \cline{2-4}
  & \textcolor[rgb]{1.00,0.00,0.00}{miR-34c}$\times$Ago2 & 0.102 & [0.033,0.118] \\
  \cline{2-4}
  & \textcolor[rgb]{1.00,0.00,0.00}{miR-34b}$\times$Ago2 & 0.054 & [0.043,0.158] \\
  \cline{2-4}
  & \textcolor[rgb]{1.00,0.00,0.00}{miR-23b}$\times$Ago2 & 0.098 & [0.036,0.136] \\
  \cline{2-4}
  & \textcolor[rgb]{1.00,0.00,0.00}{miR-24}$\times$Ago2 & 0.191 & [0.051,0.168] \\
  \cline{2-4}
  & \textcolor[rgb]{1.00,0.00,0.00}{miR-27b}$\times$Ago2 & 0.051 & [0.022,0.077] \\
  \cline{2-4}
  & miR-92$\times$Ago134 & 0.093 & [0.028,0.109] \\
  \cline{2-4}
  & miR-139$\times$Ago134 & 0.121 & [0.042,0.146] \\
  \cline{2-4}
  & miR-30d$\times$Ago134 & 0.262 & [0.047,0.145] \\
  \cline{2-4}
  & \textcolor[rgb]{1.00,0.00,0.00}{miR-34c}$\times$Ago134 & 0.121 & [0.026,0.097] \\
  \cline{2-4}
  & \textcolor[rgb]{1.00,0.00,0.00}{miR-34b}$\times$Ago134 & 0.071 & [0.032,0.137] \\
  \cline{2-4}
  & \textcolor[rgb]{1.00,0.00,0.00}{miR-23b}$\times$Ago134 & 0.081 & [0.021,0.083] \\
  \cline{2-4}
  & \textcolor[rgb]{1.00,0.00,0.00}{miR-24}$\times$Ago134 & 0.143 & [0.038,0.117] \\
  \hline
\end{tabular}
\begin{flushleft}The active credible intervals and their statistical significance computed by non-negative Bayesian LASSO for gene NOTCH1. Eight miRNAs were identified by non-negative Bayesian LASSO as targeting NOTCH1. Only the miRNAs with significance greater than $0.05$ are shown.  The miRNAs highlighted in red have been experimentally validated as down-regulating NOTCH1.
\end{flushleft}
\label{table_nbLasso_Notch}
\end{table}

\begin{table}[!ht]
\caption{\bf{The inferred statistical significance and active credible intervals using non-negative Bayesian LASSO for the MCC data.}}
\scalebox{0.64}{
\begin{tabular}{|c|c|c|c|c|c|c|c|}
  \hline
  mRNA & miRNA & Significance & Credible Interval & mRNA & miRNA & Significance & Credible Interval \\
  \hline
  {EEF1A2} & \textbf{let-7f} & \textbf{0.628} & \textbf{[0.107,0.239]} & CDKN2A& miR-125b & 0.155 & [0.051, 0.197] \\
  \cline{2-4}\cline{6-8}
   & miR-181b & 0.067 & [0.046, 0.178] &  & miR-195 & 0.057 & [0.033, 0.151] \\
  \cline{2-4}\cline{6-8}
   & miR-182 & 0.066 & [0.037, 0.151] &  & miR-29a & 0.051 & [0.066, 0.305] \\
  \cline{1-4}\cline{6-8}
  FSCN1 & \textbf{miR-133a} & \textbf{0.458} & \textbf{[0.071, 0.201]} & & \textbf{miR-99a} & \textbf{0.085} & \textbf{[0.032, 0.135]}\\
  \cline{2-8}
   & \textbf{miR-145} & \textbf{0.103} & \textbf{[0.041, 0.148]} & CEACAM5 & \textbf{mir-143} & \textbf{0.079} & \textbf{[0.058, 0.223]}\\
  \cline{2-4}\cline{6-8}
   & miR-29c & 0.291 & [0.084, 0.262] &  & \textbf{miR-145} & \textbf{0.096} & \textbf{[0.068, 0.254]}\\
  \cline{1-4}\cline{6-8}
  BRAF & let-7f & 0.053 & [0.037, 0.154] & & miR-221 & 0.057 & [0.071, 0.283]\\
  \cline{2-8}
   & \textbf{{miR-145}} & 0.150 & [0.041, 0.143] & KRT14 & \textbf{miR-143} & \textbf{0.054} & \textbf{[0.029, 0.123]}\\
  \cline{2-8}
   & miR-183 & 0.059 & [0.038, 0.157] & IGFBP6 & miR-181b & 0.115 & [0.051, 0.182] \\
  \cline{2-4}\cline{6-8}
   & \textbf{miR-192} & \textbf{0.084} & \textbf{[0.016, 0.063]} & & \textbf{miR-27a} & \textbf{0.073} & \textbf{[0.048, 0.185]} \\
  \cline{2-5}\cline{6-8}
   & miR-221 & 0.103 & [0.049, 0.188] & PIGR & \textbf{miR-125b} & \textbf{0.279} & \textbf{ [0.048, 0.152]}\\
  \cline{2-5}\cline{6-8}
   & miR-223 & 0.318 & [0.063, 0.200] &  PLK1 & let-7b & 0.314 & [0.125, 0.399] \\
  \cline{2-4}\cline{6-8}
   & miR-9 & 0.107 & [0.044, 0.167] & & \textbf{miR-100} & \textbf{0.411} & \textbf{[0.046, 0.136]} \\
  \cline{1-4}\cline{6-8}
  CCND1 & miR-130b & 0.171 & [0.176, 0.702] & & miR-99a & 0.129 & [0.031, 0.107] \\
  \cline{2-4}\cline{6-8}
   & \textbf{miR-194} & \textbf{0.097} & \textbf{[0.031, 0.138]} & & miR-99b & 0.164 & [0.075, 0.247]\\
  \cline{2-8}
   & miR-7 & 0.129 & [0.061, 0.270] & BAK1 & \textbf{miR-125b} & \textbf{0.431} & \textbf{ [0.038, 0.109]} \\
  \cline{1-4}\cline{6-8}
  HOXD10 & let-7b & 0.101 & [0.124, 0.595] & & miR-16 & 0.308 & [0.053, 0.164] \\
  \cline{2-4}\cline{6-8}
   & miR-103 & 0.050 & [0.056, 0.277] & & miR-25 & 0.117 & [0.038, 0.137]\\
  \cline{2-4}\cline{6-8}
   & miR-125b & 0.088 & [0.041, 0.203] & & miR-29a & 0.052 & [0.047, 0.194]\\
  \cline{2-4}\cline{6-8}
   & miR-141 & 0.052 & [0.026, 0.139] & & miR-32 & 0.088 & [0.032, 0.119]\\
  \cline{2-8}
   & miR-145 & 0.072 & [0.038, 0.189] & E2F1 & miR-16 & 0.059 & [0.051, 0.302]\\
  \cline{2-4}\cline{6-8}
   & miR-15b & 0.067 & [0.048, 0.230] & & \textbf{miR-195} & \textbf{0.173} & \textbf{[0.032, 0.145]}\\
  \cline{2-4}\cline{6-8}
   & miR-200b & 0.082 & [0.028, 0.134] &  & miR-23b & 0.097 & [0.044, 0.234]\\
  \cline{2-4}\cline{6-8}
   & \textbf{miR-200c} & \textbf{0.125} & \textbf{[0.028, 0.131]} & & miR-29a & 0.089 & [0.068, 0.355] \\
  \cline{2-8}
   & miR-21 & 0.054 & [0.047, 0.254] & TWIST1 & \textbf{miR-141} & \textbf{0.073} &\textbf{ [0.023, 0.113]} \\
  \cline{2-4}\cline{6-8}
   & miR-25 & 0.078 & [0.061, 0.302] & & miR-145 & 0.104 & [0.026, 0.132]\\
  \cline{2-4}\cline{6-8}
   & miR-7 & 0.078 & [0.035, 0.182] & & \textbf{miR-200a} & \textbf{0.058} & \textbf{[0.023, 0.118]}\\
  \cline{1-4}\cline{6-8}
  CDKN2A & miR-1 & 0.231 & [0.029, 0.102] & & \textbf{miR-200b} & \textbf{0.193} & \textbf{[0.028, 0.111]}\\
  \cline{2-4}\cline{6-8}
   & miR-100 & 0.087 & [0.041, 0.169] & & \textbf{miR-200c} & \textbf{0.390} & \textbf{[0.028, 0.096]}\\
  \hline
\end{tabular}
}
\begin{flushleft}Shown are the active credible intervals and their significance for the effects of miRNA targeting mRNA. Only the miRNAs with significance greater than 0.05 are listed here. The miRNAs in bold have been experimentally validated as repressing their targets.
\end{flushleft}
\label{table_nblasso_mcc}
\end{table}

\begin{table}[!ht]
\caption{\bf{The inferred statistical significance and active credible intervals using Bayesian LASSO for the MCC data.}}
\scalebox{0.64}{
\begin{tabular}{|c|c|c|c|c|c|c|c|}
  \hline
  mRNA & miRNA & Significance & Credible Interval & mRNA & miRNA & Significance & Credible Interval \\
  \hline
  {EEF1A2} & \textbf{let-7f} & \textbf{0.631} & \textbf{[0.098,0.239]} & HOXD10 & miR-7 & 0.084 & [0.026, 0.098] \\
  \hline
  FSCN1 & \textbf{miR-133a} & \textbf{0.387} & \textbf{[0.067, 0.201]} & CDKN2A & miR-1 & 0.501 & [0.039, 0.109] \\
  \cline{2-4}\cline{6-8}
   & \textbf{miR-145} & \textbf{0.088} & \textbf{[0.044, 0.175]} & & {miR-100} & {0.077} & {[0.038, 0.137]}\\
  \cline{2-4}\cline{6-8}
   & {miR-29c} & {0.252} & {[0.095, 0.331]} & & miR-125b & 0.131 & [0.048, 0.165]\\
  \cline{1-4}\cline{6-8}
  BRAF & {miR-145} & {0.120} & {[0.040, 0.138]} & & {miR-195} & {0.069} & {[0.038, 0.144]}\\
  \cline{2-4}\cline{6-8}
   & \textbf{miR-192} & \textbf{0.063} & \textbf{[0.015, 0.058]} & & \textbf{miR-99a} & \textbf{0.104} & \textbf{[0.041, 0.141]}\\
  \cline{2-5}\cline{6-8}
   & miR-221 & 0.072 & [0.043, 0.159] & CEACAM5 & \textbf{miR-143} & \textbf{0.072} & \textbf{[0.062, 0.237]}\\
  \cline{2-4}\cline{6-8}
   & miR-223 & 0.337 & [0.063, 0.194] & & \textbf{miR-145} & \textbf{0.073} & \textbf{ [0.068, 0.258]}\\
\cline{2-4}\cline{6-8}
   & miR-9 & 0.086 & [0.041, 0.151] & & miR-221 & 0.074 & [0.086, 0.323]\\
  \cline{1-4}\cline{5-8}
  CCND1 & {miR-1} & {0.181} & {[0.034, 0.117]} & IGFBP6 & {miR-181b} & {0.079} & {[0.047, 0.176]}\\
  \cline{2-4}\cline{5-8}
   & miR-130a & 0.100 & [0.063, 0.238] & PIGR & \textbf{miR-125b} & \textbf{0.238} & \textbf{[0.051, 0.170]}\\
  \cline{2-4}\cline{5-8}
   & miR-130b & 0.126 & [0.142, 0.490] & PLK1 & {let-7b} & {0.213} & {[0.101, 0.336]}\\
  \cline{2-4}\cline{6-8}
   & \textbf{miR-194} & \textbf{0.211} & \textbf{[0.033, 0.109]} & & \textbf{miR-100} & \textbf{0.533} & \textbf{[0.048, 0.132]}\\
  \cline{2-4}\cline{6-8}
   & miR-7 & 0.166 & [0.054, 0.180] & & miR-99a & 0.161 & [0.076, 0.239]\\
  \cline{1-4}\cline{6-8}
  HOXD10 & {let-7b} & {0.075} & {[0.097, 0.362]} & & {miR-99b} & {0.161} & { [0.076, 0.239]}\\
  \cline{2-4}\cline{5-8}
   & miR-103 & 0.080 & [0.058, 0.216] & BAK1 & \textbf{miR-125b} & \textbf{0.467} & \textbf{[0.037, 0.106]}\\
  \cline{2-4}\cline{6-8}
   & let-107 & 0.064 & [0.054, 0.205] & & miR-16 & 0.378 & [0.056, 0.168] \\
  \cline{2-4}\cline{6-8}
   & miR-125b & 0.110 & [0.036, 0.128] & & miR-25 & 0.087 & [0.035, 0.129]\\
  \cline{2-4}\cline{6-8}
   & miR-141 & 0.071 & [0.024, 0.092] & & miR-32 & 0.070 & [0.029, 0.112]\\
  \cline{2-4}\cline{5-8}
   & miR-145 & 0.087 & [0.034, 0.122] & E2F1 & miR-16 & 0.057 & [0.042, 0.160] \\
  \cline{2-4}\cline{6-8}
   & miR-15b & 0.093 & [0.050, 0.179] & & \textbf{miR-195} & \textbf{0.58} & \textbf{[0.048, 0.129]}\\
  \cline{2-4}\cline{6-8}
   & miR-195 & 0.050 & [0.029, 0.118] & & miR-23b & 0.091 & [0.039, 0.140]\\
  \cline{2-4}\cline{6-8}
   & miR-200b & 0.098 & [0.026, 0.095] & & miR-29a & 0.067 & [0.051, 0.190]\\
  \cline{2-4}\cline{5-8}
   & \textbf{miR-200c} & \textbf{0.219} & \textbf{[0.029, 0.099]} & TWIST1 & \textbf{miR-200c} & \textbf{0.691} &\textbf{ [0.032, 0.086]} \\
  \cline{2-4}\cline{6-8}
   & miR-21 & 0.057 & [0.034, 0.130] && {miR-145} & {0.104} &{ [0.018, 0.068]}\\
  \cline{2-4}\cline{5-8}
   & miR-25 & 0.099 & [0.065, 0.232] && \textbf{miR-200b} & \textbf{0.153} & \textbf{ [0.028, 0.083]}\\
  \hline

   \end{tabular}
}
\begin{flushleft}Shown are the active credible intervals and their significance for the effects of miRNA targeting mRNA. Only the miRNAs with significance greater than 0.05 are listed here. The miRNAs in bold have been experimentally validated as repressing their targets.
\end{flushleft}
\label{table_mcc1}
\end{table}

\begin{table}[!ht]
\caption{\bf{{The number of estimated interactions which have been experimentally validated.}}}
\scalebox{0.9}{
\begin{tabular}{|c|c|c|c|c|c|c|c|c|}
  \hline
  Data & No. of Validated Targets &  LASSO &  BLASSO & nBLASSO & \multicolumn{3}{c}{nLASSO} \\
  &  & $\beta=0$ & $\alpha=0.05$ & $\alpha=0.05$ & $\beta=0.0001$ & $\beta=0.001$ & $\beta=0.01$\\ 
  \hline
  MCC & 23 & 15 &  17 & 20 & 20& 18& 17\\
  \hline
  Broad & 109 & 66 &  75 & 79 & 75 & 70& 65\\
  \hline
  Madison & 153  & 81 &  116 & 138 &60& 30 &3 \\
  \hline
  MSKCC & 116 & 84 &  95 &98 & 98& 69&55  \\
  \hline
\end{tabular}
}
\begin{flushleft}{Shown are the number of estimated interactions by using the LASSO, nLASSO, BLASSO and nBLASSO. Only the experimentally validated interactions were counted. Various values of the threshold $\boldsymbol{\beta}$ were used for the point estimation methods to select the estimated interactions. For the Bayesian methods, a significance level $\alpha=0.05$ was used to select the inferred interactions.}
\end{flushleft}
\label{final_res}
\end{table}

\clearpage

%
%

\begin{table}[!ht]
\caption{\bf{The areas under the ROC curves for the false positive range $[0,0.1]$.}}

\begin{tabular}{|c|c|c|c|c|}
  \hline
  AUC & Broad & Madison & MCC & MSKCC \\
  \hline
  LSR & 0.0 & 0.0 & 0.0084 & 0.0 \\
  \hline
  Ridge & 0.0036 & 0.0151 & 0.0229 & 0.0101 \\
  \hline
  LASSO & 0.0025 & 0.0110 & 0.0165 & 0.0073 \\
  \hline
  BLASSO & 0.0264 & 0.0114 & 0.0222 & 0.0214 \\
  \hline
  nLASSO & 0.0303 & 0.0211 & 0.0338 & 0.0311 \\
  \hline
  nBLASSO & 0.0274 & 0.0178 & 0.0321 & 0.0260 \\
  \hline
\end{tabular}

\label{auc}
\end{table}

\end{document}